\documentclass[11pt]{article}
\usepackage{amssymb}
\usepackage{colortbl}
\usepackage{amsfonts,amsmath, longtable}

        \def\theequation{\thesection.\arabic{equation}}

\newcommand{\tr}{{\rm tr}}
\newcommand{\ti}[1]{\tilde{#1}}

\newcommand{\mL}{{\mathcal L}}

\newcommand{\Mat}{ {\rm Mat}(N,\mathbb C) }

\newcommand{\mC}{\mathbb C}

\newtheorem{predl}{Proposition}[section]
\newtheorem{lemma}{Lemma}[section]

\def\beq{\begin{equation}}
\def\eq{\end{equation}}
\def\p{\partial}

\newtheorem{theor}{Theorem}

\def\res{\mathop{\hbox{Res}}\limits}

\begin{document}

\setcounter{page}{1}

\begin{center}

\

\vspace{-10mm}

{\Large{\bf{Higher rank generalization of 11-vertex rational R-matrix:}}}

\vspace{3mm}

{\Large{\bf{IRF-Vertex relations and associative Yang-Baxter equation }}}

\vspace{3mm}



 \vspace{15mm}

 {\Large {K. Atalikov}}$\,^{\diamond\,\bullet}$
\qquad\quad\quad
 {\Large {A. Zotov}}$\,^{\diamond\,\bullet}$

  \vspace{5mm}

$\diamond$ -- {\em Steklov Mathematical Institute of Russian
Academy of Sciences,\\ Gubkina str. 8, 119991, Moscow, Russia}


$\bullet$ -- {\em NRC ''Kurchatov Institute'',\\
Kurchatova sq. 1, 123182, Moscow, Russia}


   \vspace{3mm}

 {\small\rm {e-mails: kantemir.atalikov@yandex.ru, zotov@mi-ras.ru}}

\end{center}

\vspace{0mm}

\begin{abstract}
We study ${\rm GL}_N$ rational $R$-matrix, which turns into the 11-vertex $R$-matrix in the $N=2$ case. First, we describe its relations to dynamical and semi-dynamical $R$-matrices using the IRF-Vertex type transformations. As a by-product a new explicit form for ${\rm GL}_N$ $R$-matrix is derived. Next, we prove the quantum and the associative Yang-Baxter equations.
A set of other $R$-matrix properties and $R$-matrix identities are proved as well.
\end{abstract}

\bigskip

{\small{
\tableofcontents
}}


\newpage

\section{Introduction}
\setcounter{equation}{0}
\paragraph{$R$-matrix.} Quantum $R$-matrix is a solution of the quantum Yang-Baxter equation \cite{Yang,Baxter,Sklyanin}:
 \beq\label{w01}
 \begin{array}{c}
  \displaystyle{
 R_{12}^\hbar R_{13}^\hbar R_{23}^\hbar
=R_{23}^\hbar R_{13}^\hbar R_{12}^\hbar\,,
   \qquad R^\hbar_{ab} = R^\hbar_{ab}(z_a-z_b)\,.
  }
 \end{array}
\eq
It is a fundamental object in studies of quantum exactly solvable models based on RTT relations and Bethe ansatz technique. It also appears in different areas of mathematics, theoretical and mathematical physics.
In this paper we deal with $R$-matrices in the fundamental representation of the ${\rm GL}(N,\mC)$ Lie group, i.e.
$R$-matrix $R^\hbar_{12}(z)$ is a matrix-valued function in $\Mat^{\otimes 2}$ of two complex variables
($\hbar$ is the Planck constant and $z$ is the spectral parameter). It has the form
 \beq\label{w02}
 \begin{array}{c}
  \displaystyle{
 R_{12}^\hbar(z)=\sum\limits_{i,j,k,l=1}^N R_{ij,kl}(\hbar,z)\, E_{ij}\otimes E_{kl}\,,
  }
 \end{array}
\eq
where the set of matrices $E_{ij}$ is the standard matrix basis in $\Mat$: $(E_{ij})_{ab}=\delta_{ia}\delta_{jb}$.
The indices $12$ mean the numbers of tensor components, where $R$-matrix acts nontrivially as a linear operator.
Equation (\ref{w01}) is written in $\Mat^{\otimes 3}$. Then
 \beq\label{w03}
 \begin{array}{c}
  \displaystyle{
  R_{12}^\hbar(z)=\sum\limits_{i,j,k,l=1}^N R_{ij,kl}(\hbar,z)\, 1_N\otimes E_{ij}\otimes E_{kl}\,,\qquad
 R_{13}^\hbar(z)=\sum\limits_{i,j,k,l=1}^N R_{ij,kl}(\hbar,z)\, E_{ij}\otimes 1_N\otimes E_{kl}\,,
  }
 \end{array}
\eq
and similarly for $R_{23}^\hbar(z)$, where $1_N$ is $N\times N$ identity matrix. Permutation of indices $12\rightarrow 21$ means the action of the permutation operator $P_{12}$:
 \beq\label{w04}
 \begin{array}{c}
  \displaystyle{
 R_{21}^\hbar(z)=\sum\limits_{i,j,k,l=1}^N R_{ij,kl}(\hbar,z)\, E_{kl}\otimes E_{ij}=P_{12}R_{12}^\hbar(z)P_{12}\,,
 \qquad P_{12}=\sum\limits_{i,j=1}^N E_{ij}\otimes E_{ji}\,.
  }
 \end{array}
\eq

\paragraph{Properties and normalization.} The quantum Yang-Baxter equation (\ref{w01}) provides a set of equations for the coefficient functions $R_{ij,kl}(\hbar,z)$. Notice that (\ref{w01}) defines $R$-matrix up to multiplication by an arbitrary function.
This ambiguity can be fixed in different ways depending on additional properties of $R$-matrix. In this paper we consider $R$-matrices satisfying also the {\em unitarity property}
 \beq\label{w05}
 \begin{array}{c}
  \displaystyle{
 R_{12}^\hbar(z)R_{21}^\hbar(-z)=f(\hbar,z)\,1_N\otimes 1_N
  }
 \end{array}
\eq
(where $f(\hbar,z)$ is some function) and the {\em skew-symmetry property}
 \beq\label{w06}
 \begin{array}{c}
  \displaystyle{
 R_{12}^\hbar(z)=-R_{21}^{-\hbar}(-z)\,.
  }
 \end{array}
\eq
For unitary $R$-matrices a custom choice of the function $f(\hbar,z)$ is $1$ but we fix normalization in
a different way\footnote{It follows from (\ref{w05}) and (\ref{w07}) that
 ${\ti R}_{12}^\hbar(z){\ti R}_{21}^\hbar(-z)=1_N\otimes 1_N$ for ${\ti R}_{12}^\hbar(z)=R_{12}^\hbar(z)/\phi(\hbar,z)$.}:
 \beq\label{w07}
 \begin{array}{c}
  \displaystyle{
 f(\hbar,z)=\frac{1}{\hbar^2}-\frac{1}{z^2}=\phi(\hbar,z)\phi(\hbar,-z)\,,\qquad
 \phi(\hbar,z)=\frac{1}{\hbar}+\frac{1}{z}\,.
  }
 \end{array}
\eq
Hence, the function $\phi(\hbar,z)$ can be considered as ${\rm GL}_1$ $R$-matrix, or put it differently,
${\rm GL}_N$ $R$-matrix is a matrix analogue of function $\phi(\hbar,z)$.

Another important property is the classical limit. It is an expansion of $R$-matrix near $\hbar=0$:
 \beq\label{w08}
 \begin{array}{c}
  \displaystyle{
 R^{\hbar}_{12} (z)=\frac{1}{\hbar}\, 1_N \otimes 1_N + r_{12}(z) + \hbar \: m_{12} (z) + O(\hbar^2)\,,
  }
 \end{array}
\eq
The first nontrivial coefficient of the expansion is the classical $r$-matrix $r_{12}(z)$. From (\ref{w01}) one easily obtains the classical Yang-Baxter equation:
 \beq\label{w09}
 \begin{array}{c}
  \displaystyle{
[r_{12}(z_1 - z_2), r_{13}(z_1 - z_3)]+[r_{12}(z_1 - z_2), r_{23}(z_2 - z_3)]+[r_{13}(z_1 - z_2), r_{23}(z_2 - z_3)]=0\,.
  }
 \end{array}
\eq
The local behaviour of $R_{12}^\hbar(z)$ near $z=0$ is defined by the condition:
 \beq\label{w10}
 \begin{array}{c}
  \displaystyle{
\res\limits_{z=0}R_{12}^\hbar(z)=P_{12}\,.
  }
 \end{array}
\eq

\paragraph{Yang's $R$-matrix and its deformation.} The simplest example of $R$-matrix satisfying all above mentioned properties is given by the rational  Yang's $R$-matrix:
 \beq\label{w11}
 \begin{array}{c}
  \displaystyle{
R_{12}^{{\rm Yang},\hbar}(z)=\frac{1_N\otimes 1_N}{\hbar}+\frac{P_{12}}{z}\,.
  }
 \end{array}
\eq
In $N=2$ case it is the widely known 6-vertex XXX $R$-matrix normalized as in (\ref{w07}):
 \beq\label{w12}
 \begin{array}{c}
  \displaystyle{
 R_{12}^{{\rm 6v},\hbar}(z)=
  \left(
  \begin{array}{cccc}
  1/\hbar+1/z & 0 & 0 & 0
  \\
  0 & 1/\hbar & 1/z & 0
  \\
   0 & 1/z & 1/\hbar & 0
   \\
   0 & 0 & 0 & 1/\hbar+1/z
   \end{array}
   \right)\,.
  }
 \end{array}
\eq
It was mentioned by I. Cherednik in \cite{Chered} that ${\rm GL}_2$ $R$-matrix (\ref{w12}) has the following
 11-vertex deformation:
 \beq\label{w13}
 \begin{array}{c}
  \displaystyle{
 R_{12}^{{\rm 11v},\hbar}(z)=
  \left(
  \begin{array}{cccc}
  1/\hbar+1/z & 0 & 0 & 0
  \\
  -z-\hbar & 1/\hbar & 1/z & 0
  \\
   -z-\hbar & 1/z & 1/\hbar & 0
   \\
   -z^3-\hbar^3-2z^2\hbar-2z\hbar^2 & z+\hbar & z+\hbar & 1/\hbar+1/z
   \end{array}
   \right)\,.
  }
 \end{array}
\eq
The latter $R$-matrix also satisfies all above mentioned equations and properties. The corresponding quantum integrable spin chains were discussed in \cite{Manoj,LOZ142}. It was shown in \cite{Smirnov} that the 11-vertex $R$-matrix appears from the elliptic Baxter's $R$-matrix through a special limiting procedure. An algorithm for computation of higher rank generalizations of (\ref{w13}) was suggested as well based on the limiting procedure applied to ${\rm GL}_N$ elliptic Baxter-Belavin $R$-matrix. Explicit form for $R$-matrix in ${\rm GL}_N$ case was derived
in \cite{LOZ14} via classical analogue of IRF-Vertex transformation relating the rational $N$-body Ruijsenaars-Schneider model
and  relativistic top for ${\rm GL}_N$ Lie group. In a similar way the corresponding classical $r$-matrix was evaluated in \cite{AASZ} using gauge equivalence of the rational Calogero-Moser model and a certain
integrable rational top-like model. Explicit formulae obtained in \cite{AASZ,LOZ14} are quite complicated. See Appendix for details.

The 11-vertex $R$-matrix (as well as its higher rank generalization) can be viewed as deformation of the Yang's $R$-matrix in the following sense:
 \beq\label{w131}
 \begin{array}{c}
  \displaystyle{
 \lim\limits_{\epsilon\rightarrow 0}\epsilon R_{12}^{{\rm 11v},\hbar\epsilon}(z\epsilon)=R_{12}^{{\rm Yang},\hbar}(z)\,.
  }
 \end{array}
\eq

Notice that up till now we discussed $R$-matrices depending on the difference of spectral parameters $R^\hbar_{ab}(z_a,z_b)=R^\hbar_{ab}(z_a-z_b)$. All equations and conditions can be extended to the case
$R^\hbar_{ab}(z_a,z_b)\neq R^\hbar_{ab}(z_a-z_b)$. In this form the answer for ${\rm GL}_N$ rational $R$-matrix
was obtained in \cite{BK} by studying vector bundles on the cuspidal cubic curve. Presumably, the answer from \cite{BK} is gauge equivalent to $R^\hbar_{ab}(z_a-z_b)$ from \cite{Smirnov,LOZ14}.

\paragraph{Dynamical $R$-matrices.} All above mentioned $R$-matrices are of vertex type.
Another wide class of $R$-matrices is of IRF type \cite{Baxter2,Jimbo}. The corresponding $R$-matrices are called dynamical (vertex type $R$-matrices are also called non-dynamical) since they depend on additional (dynamical) parameters $q_1,...,q_N$. They satisfy the quantum dynamical Yang-Baxter
equation (or the Gervais-Neveu-Felder equation)
\cite{Felder}:
  \beq\label{w15}
  \begin{array}{c}
  \displaystyle{
 R^\hbar_{12}(z_1-z_2|\,q-\hbar^{(3)})R^\hbar_{13}(z_1-z_3|\,q)R^\hbar_{23}(z_2-z_3|\,q-\hbar^{(1)})=\hspace{40mm}
}
\\ \ \\
  \displaystyle{
\hspace{40mm}
=R^\hbar_{23}(z_2-z_3|\,q)R^\hbar_{13}(z_1-z_3|\,q-\hbar^{(2)})R^\hbar_{12}(z_1-z_2|\,q)\,,
 }
 \end{array}
 \eq
where the shifts of dynamical arguments $q$ are performed by the following rule:
  \beq\label{w16}
  \begin{array}{c}
  \displaystyle{
R^\hbar_{12}(z_1,z_2|\,q+\hbar^{(3)})=P_3^\hbar\,
R^\hbar_{12}(z_1,z_2|\,q)\, P_3^{-\hbar} \,,\quad
P_3^\hbar=\sum\limits_{k=1}^N 1\otimes 1\otimes E_{kk}
\exp(N\hbar\frac{\p}{\p q_k})\,.
 }
 \end{array}
 \eq
 The factor $N$ in the shift operator is due to our special choice of normalization of variables.

 Certain dynamical and non-dynamical $R$-matrices are related by the so-called IRF-Vertex transformation \cite{Baxter2,Jimbo}:
  \beq\label{w17}
 \begin{array}{c}
  \displaystyle{
R^{\hbox{\tiny{Vertex}},\hbar}_{12}(z_1-z_2)=
}
\\ \ \\
  \displaystyle{
  =
g_2(z_2,q)\,
g_1(z_1,q-\hbar^{(2)})\,R^{\hbox{\tiny{Dynam}}}_{12}(\hbar,z_1-z_2|\,q)
g^{-1}_2(z_2,q-\hbar^{(1)})g_1^{-1}(z_1,q)\,,
 }
 \end{array}
 \eq
 where $g(z,q)\in\Mat$ is some special matrix providing the IRF-Vertex transformation and
  \beq\label{w171}
  \begin{array}{c}
  \displaystyle{
  g_1(z_1,q)=g(z_1,q)\otimes 1_N\,,\qquad
    g_2(z_2,q)=1_N\otimes g(z_2,q)\,,
    }
\\ \ \\
  \displaystyle{
g_1(z_1,q-\hbar^{(2)})=P_2^{-\hbar}g_1(z_1,q)P_2^\hbar\,,\qquad
g_2(z_2,q-\hbar^{(1)})=P_1^{-\hbar}g_2(z_2,q)P_1^\hbar
 }
 \end{array}
 \eq
 with $P_i^\hbar$ defined in (\ref{w16}).
  Relation (\ref{w17}) can be considered as a quantum gauge transformation. It is highly non-trivial. The r.h.s. is independent of $q_1,...,q_N$, while
 all matrix factors in the r.h.s. depend on the dynamical variables. Also, the matrices of gauge transformation depend on either $z_1$ or $z_2$ but the result depend on the difference of spectral parameters $z_1-z_2$ only.

\paragraph{Semi-dynamical $R$-matrices.} Another interesting class of $R$-matrices was introduced by G. Arutyunov, L. Chekhov and S. Frolov in \cite{Arut}, see also \cite{Avan1,SeZ}. These $R$-matrices (they were called semi-dynamical) also depend on the dynamical parameters $q_1,...,q_N$
 but the Yang-Baxter equation is different:
  \beq\label{w18}
  \begin{array}{c}
  \displaystyle{
 R^\hbar_{12}(z_1,z_2|\,q)R^\hbar_{13}(z_1-\hbar,z_3-\hbar|\,q)R^\hbar_{23}(z_2,z_3|\,q)=\hspace{50mm}
}
\\ \ \\
  \displaystyle{
\hspace{40mm}
=R^\hbar_{23}(z_2-\hbar,z_3-\hbar|\,q)R^\hbar_{13}(z_1,z_3|\,q)R^\hbar_{12}(z_1-\hbar,z_2-\hbar|\,q)\,.
 }
 \end{array}
 \eq
 An analogue of the IRF-Vertex relation (\ref{w17}) is quite simple in this case (see \cite{SeZ}):
 \beq\label{w19}
  \begin{array}{c}
  \displaystyle{
 R^{\hbox{\tiny{Vertex}},\hbar}_{12}(z_1-z_2)=
 }
\\ \ \\
  \displaystyle{
 =g_2(z_2,q)\,
 g_1(z_1+\hbar,q)\,
 R^{\hbox{\tiny{Semi-dynam}}}_{12}(\hbar,z_1,z_2|\,q)\, g_2^{-1}(z_2+\hbar,q)
 g_1^{-1}(z_1,q)\,.
 }
 \end{array}
 \eq
 Notice that in contrast to dynamical $R$-matrix, the semi-dynamical one does not depend on the difference of spectral parameters\footnote{Of course one can make a dynamical $R$-matrix also not to depend on the difference of spectral parameters by applying some gauge transformation. Here we mean that dynamical $R$-matrices can be chosen in a gauge, where they depend on the difference of spectral parameters. It is not true for semi-dynamical $R$-matrices.}.

 By comparing the right hand sides of (\ref{w17}) and (\ref{w19}) we conclude
  \beq\label{w191}
  \begin{array}{c}
  \displaystyle{
 R^{\hbox{\tiny{Semi-dynam}}}_{12}(\hbar,z_1,z_2|\,q)=
 }
\\ \ \\
  \displaystyle{
 =
 g_1^{-1}(z_1+\hbar,q)\,g_1(z_1,q-\hbar^{(2)})\,
 R^{\hbox{\tiny{Dynam}}}_{12}(\hbar,z_1-z_2|\,q)\, g^{-1}_2(z_2,q-\hbar^{(1)})\, g_2(z_2+\hbar,q)\,.
 }
 \end{array}
 \eq

\paragraph{Associative
Yang-Baxter equation.} A certain class of  $R$-matrices of vertex type satisfies the quadratic relation known as
the associative
Yang-Baxter equation \cite{FK}:
 \beq\label{w14}
 \begin{array}{c}
  \displaystyle{
 R^{\hbar}_{12} R^{\eta}_{23} = R^{\eta}_{13} R^{\hbar-\eta}_{12} + R^{\eta-\hbar}_{23} R^{\hbar}_{13}\,,
   \qquad R^x_{ab} = R^x_{ab}(z_a-z_b)\,.
  }
 \end{array}
\eq
More precisely, solutions of the quantum Yang-Baxter equation do not always satisfy (\ref{w14}). Vice versa,
not all solutions of the associative Yang-Baxter equation satisfy (\ref{w01}). However, if we consider solutions of
(\ref{w14}) with additional properties (\ref{w05})-(\ref{w06}) then such a linear operator $R_{12}^\hbar(z)$ do satisfy the quantum Yang-Baxter equation (\ref{w01}), so that it is indeed a quantum $R$-matrix.
A simple proof of this fact can be found in \cite{LOZ141}.
To summarize,
a skew-symmetric and unitary solution of (\ref{w14}) in the fundamental representation of ${\rm GL}_N$ Lie group
is a quantum $R$-matrix. 

The rational $R$-matrix found in \cite{BK} (where $R^\hbar_{ab}(z_a,z_b)\neq R^\hbar_{ab}(z_a-z_b)$) by construction also satisfies analogue
of (\ref{w14}) for $R$-matrices which do not depend on the difference of the spectral parameters. Let us also mention
that a certain set of rational $R$-matrices satisfies the non-homogeneous associative Yang-Baxter equation \cite{OgP}.

To our best knowledge an analogue of the associative Yang-Baxter equation for dynamical $R$-matrices is unknown.
However, it is known for semi-dynamical $R$-matrices \cite{SeZ}:
  \beq\label{w20}
  \begin{array}{c}
  \displaystyle{
 R^\hbar_{12}(z_1+\eta,z_2+\eta)
 R^{\eta}_{23}(z_2+\hbar,z_3+\hbar)=
 }
 \\ \ \\
  \displaystyle{
 =R^{\eta}_{13}(z_1+\hbar,z_3+\hbar)R_{12}^{\hbar-\eta}(z_1+\eta,z_2+\eta)
 +R^{\eta-\hbar}_{23}(z_2+\hbar,z_3+\hbar)R^\hbar_{13}(z_1+\eta,z_3+\eta)\,,
 }
 \end{array}
 \eq
where $R^\hbar_{ab}(z_1,z_2)=R^{\hbox{\tiny{Semi-dynam}}}_{ab}(\hbar,z_1,z_2|\,q)$.

\paragraph{Purpose of the paper.}
In this paper we discuss the 11-vertex $R$-matrix (\ref{w13}) and its higher rank generalization.
In Section 2 we review the properties of the matrix $g(z,q)$ entering (\ref{w17}), (\ref{w19}) in the rational case
based on results of \cite{AASZ}. The IRF-Vertex relations (\ref{w17}), (\ref{w19}), (\ref{w191}) are known in the elliptic and (partly) trigonometric cases but are not known for the rational case.
In Section 3 we consider dynamical and semi-dynamical rational $R$-matrices and prove relation (\ref{w191}). By applying the IRF-Vertex transformation
(\ref{w19}) we derive the rational ${\rm GL}_N$ $R$-matrix of vertex type in a new form.
This new form allows to prove a set of important properties.
In Section 4
we prove associative Yang-Baxter equation (\ref{w14}) for the rational vertex-type $R$-matrix.
Unitarity (\ref{w05}), skew-symmetry (\ref{w06}) and the symmetry of arguments are also proved.

Our study is motivated by different applications of such $R$-matrices to construction of classical and quantum integrable systems \cite{TrZ,MZ} including 1+1 field theories \cite{AtZ}.

\section{Rational IRF-Vertex transformation matrix}
\setcounter{equation}{0}

Let $q_1,...,q_N$ be a set of $\mC$-valued variables and
 \beq\label{e1610}
 \begin{array}{c}
  \displaystyle{
{\bar q}_j=q_{j}-\frac1N\sum\limits_{k=1}^N q_k\,,\qquad
\sum\limits_{k=1}^N {\bar q}_k=0\,.
  }
 \end{array}
\eq
Following \cite{AASZ} introduce the matrix $g(z)\in\Mat$:
 \beq\label{e161}
 \begin{array}{c}
  \displaystyle{
 g(z)=g(z,q_1,...,q_N)=\Xi(z,{ q})D^{-1}(q)\,,\qquad \Xi(z,{ q})\,,D(q)\in\Mat\,,
  }
 \end{array}
\eq
where
   \begin{equation}
   \label{e162}
   \begin{array}{c}
   \Xi_{ij}(z,{q}) =(z+{\bar q}_{j})^{\varrho(i)}\,,
     \qquad
   D_{ij}({q}) =\delta_{ij}\prod\limits_{k \neq i}^{N}(q_{i}-q_{k})
   \end{array}
   \end{equation}
   with
\begin{equation}
   \label{e163}
   \varrho(i) = \left\{\begin{array}{lll}
   i-1 &  \hbox{for} & 1 \leq i \leq N-1,\\
   N &  \hbox{for} & i=N\,,
   \end{array}\right.
   \end{equation}
   so that
 \beq\label{e164}
 \begin{array}{c}
\Xi(z, { q})=\left(\begin{array}{llll}
1 & 1 & \ldots & 1 \\
z+{\bar q}_{1} & z+{\bar q}_{2} & \ldots & z+{\bar q}_{N} \\
\vdots & \vdots & \vdots & \vdots \\
\left(z+{\bar q}_{1}\right)^{N-2} & \left(z+{\bar q}_{2}\right)^{N-2} & \ldots & \left(z+{\bar q}_{N}\right)^{N-2} \\
\left(z+{\bar q}_{1}\right)^{N} & \left(z+{\bar q}_{2}\right)^{N} & \ldots & \left(z+{\bar q}_{N}\right)^{N}
\end{array}\right)\,.
 \end{array}
\eq
The following properties of the defined above matrix $\Xi$ hold:

1) {\em Determinant}:
 \beq\label{e166}
 \begin{array}{c}
  \displaystyle{
 \det\Xi(z,q)=Nz\prod\limits_{i>j}^N(q_i-q_j)
  }
 \end{array}
\eq
and for $\Xi_{ij}(x)=x_j^{\rho(i)}$ we have
 \beq\label{e1661}
 \begin{array}{c}
  \displaystyle{
 \det\Xi(x)=\Big(\sum\limits_{k=1}^N x_k\Big)\prod\limits_{i>j}^N(x_i-x_j)\,.
  }
 \end{array}
\eq

2) {\em Inverse} of $\Xi(z,q)$. Define the set of elementary symmetric functions of $N$ variables $x_1,...,x_N$:
\beq\label{e167}
 \begin{array}{c}
 \displaystyle{
\prod\limits_{k=1}^{N} \,(\zeta-x_{k})=
\sum\limits_{k=0}^{N} (-1)^{k} \zeta^{k} \sigma_{k}(x_{1},...,x_{N})\,,
}
\\ \ \\
 \displaystyle{
\sigma_{N-d}(x)=(-1)^N\sum\limits_{1 \leq i_{1} <
i_{2}...<i_{d}\leq N} x_{i_{1}} x_{i_{2}}...x_{i_{d}}\,,\ \ \
d=0,...,N\,.
 }
 \end{array}
 \eq
 Similarly, define $N$ sets of functions
 $\stackrel{k}{\sigma}_s(x)$, $k=1,...,N$:
 \beq\label{e1671}
 \begin{array}{c}
 \displaystyle{
-\prod\limits_{m\neq k}^{N} \,(\zeta-x_{m})=\p_{x_k} \prod\limits_{k=1}^{N} \,(\zeta-x_{k})=\sum\limits_{s=0}^{N-1} (-1)^{s} \zeta^{s}
\stackrel{k}{\sigma}_s(x)\,.
 }
 \end{array}
 \eq
 The latter functions naturally appear in the inverse of Vandermonde matrix $V_{ij}(x)=x_j^{i-1}$ and similarly in the inverse of $\Xi_{ij}(x)=x_j^{\rho(i)}$:
  \beq\label{e1672}
 \begin{array}{c}
 \displaystyle{
V^{-1}_{kj}(x)=\frac{(-1)^j\stackrel{k}{\sigma}_{j-1}(x)}{\prod\limits_{s:s\neq
k}^{N} \,(x_k-x_{s})}\,,
\qquad
\Xi^{-1}_{k
j}({x})=(-1)^{\varrho(j)}\,\frac{\stackrel{k}{\sigma}_{\varrho(j)-1}(x)
 -(\sum\limits_{s:s\neq k}^N x_s) \stackrel{k}{\sigma}_{\varrho(j)}(x)}{(\sum\limits_{s=1}^{N}
\,x_{s})\prod\limits_{s:s\neq k}^N
(x_{k}-x_{s})}
 }
 \end{array}
 \eq
 or
  \beq\label{e16721}
 \begin{array}{c}
 \displaystyle{
\Xi^{-1}_{k
j}({x})=(-1)^{\varrho(j)}\,\frac{\sigma_{\varrho(j)}({x})}{(\sum\limits_{s=1}^{N}
\,x_{s})\prod\limits_{s\neq k}^N
(x_{k}-x_{s})}-(-1)^{\varrho(j)}\,\frac{\stackrel{k}{\sigma}_{\varrho(j)}({x})}{\prod\limits_{s\neq
k}^{N} (x_{k}-x_{s})} }\,.
 \end{array}
 \eq
 Finally, plugging $x_j=z+{\bar q}_j$ into (\ref{e1672}) we get
   \beq\label{e16722}
 \begin{array}{c}
 \displaystyle{
g^{-1}_{kj}({z,q})=(-1)^{\varrho(j)}\Big(\frac{\sigma_{\varrho(j)}({x})}{Nz}\,-
\stackrel{k}{\sigma}_{\varrho(j)}({x})\Big)\,,\qquad x_j=z+q_j-\frac1N\sum\limits_{k=1}^N q_k\,.
}
 \end{array}
 \eq
 It is also possible to decompose the expression for $g^{-1}(z,q)$ in powers of $z$:
  \beq\label{e1673}
 \begin{array}{c}
 \displaystyle{
 \Xi^{-1}_{mj}(z,{
 q})=\frac{(-1)^{\varrho(j)}}{Nz}\left\{
 \sigma_{\varrho(j)}({{ {\bar q} }})+\sum\limits_{s=1}^{N-j} z^s\left[
 \sigma_{s+j-1}({ {\bar q} }) \left(\!\begin{array}{c} s+j-1\\
 j-1\end{array}\!\right) \right.
 \right.
 }
\\
\
\\
\displaystyle{ \left.\left.
-N \stackrel{m}{\sigma}_{s+j-2}\!({{ {\bar q} }}) \left(\!\begin{array}{c} s+j-2\\
 j-1\end{array}\!\right)
 \right] -(N-j)\,z^{N-j+1} \stackrel{m}{\sigma}_{N-1}\!({{ {\bar q} }})  \left(\!\begin{array}{c} N\\
 j-1\end{array}\!\right) \right\}\,.
 }
 \end{array}
 \eq
Details can be found in the Appendix of \cite{AASZ}.

3) {\em Degeneration at $z=0$}. It follows from (\ref{e166}) that $g(z,q)$ is degenerated at $z=0$. Therefore, the matrix $g(0,q)$ should have a nontrivial kernel. This kernel is one dimensional. It is generated by the column-vector
$a$ with all entries equal to 1:
 \beq\label{e16733}
 \begin{array}{c}
 \displaystyle{
g(0,q)a=0\,,\qquad a=(1,...,1)^T\,. 
 }
 \end{array}
 \eq

4) {\em Factorization}. Introduce the matrix
 \beq\label{e1674}
 \begin{array}{c}
 \displaystyle{
\mL^\eta_{ij}(z)=\eta\Big(\frac{1}{q_i-q_j+\eta}-\frac{1}{Nz}\Big)\prod\limits_{k:k\neq j}^N\frac{q_j-q_k-\eta}{q_j-q_k}\,.
 }
 \end{array}
 \eq
Then it is represented in the factorized  form:
 \beq\label{e1675}
 \begin{array}{c}
 \displaystyle{
\mL^\eta(z)=g^{-1}(z,q)g(z-\eta,q)\,.
 }
 \end{array}
 \eq
In fact, a more general relation holds, which is similar to factorization of the Cauchy matrix through Vandermonde matrices. Consider $C(z)\in\Mat$
 \beq\label{e1676}
 \begin{array}{c}
 \displaystyle{
 C_{ij}(z)=\Big(\frac{1}{{\bar q}_i-{\bar u}_j+\eta}-\frac{1}{Nz}\Big)
 \frac{\prod\limits_{k=1}^N({\bar u}_j-{\bar q}_k-\eta)}{\prod\limits^N_{k:k\neq i}({\bar q}_i-{\bar q}_k)}\,,
 }
 \end{array}
 \eq
where the set of variables ${\bar u}_k$ satisfies ${\bar u}_1+...+{\bar u}_N=0$ similarly to (\ref{e1610}). Then
 \beq\label{e1677}
 \begin{array}{c}
 \displaystyle{
 C(z)=-\Xi^{-1}(z, q)\Xi(z-\eta, u)\,.
 }
 \end{array}
 \eq
Discussion of the factorization properties and their geometrical meaning can be found in \cite{VZ}.

\section{IRF-Vertex relations and higher rank generalizations of 11-vertex $R$-matrix}
\setcounter{equation}{0}

\subsection{Dynamical and semi-dynamical $R$-matrices}
The dynamical ${\rm GL}_N$ rational $R$-matrix is as follows:
 \beq\label{w31}
 \begin{array}{c}
 \displaystyle{
 R^{\hbox{\tiny{Dynam}}}_{12}(\hbar,z_1-z_2|\,q)=
 \sum\limits_{i\neq j}^N \Big( \frac{1}{z_1-z_2}+\frac{N}{q_j-q_i} \Big)E_{ij}\otimes E_{ji}
+ \sum\limits_{i\neq j}^N \Big( \frac{1}{\hbar}+\frac{N}{q_i-q_j} \Big)E_{ii}\otimes E_{jj}+
 }
\\
 \displaystyle{
 +\Big(\frac{1}{z_1-z_2}+\frac{1}{\hbar}\Big) \sum\limits_{i=1}^N E_{ii}\otimes E_{ii}\,.
 }
 \end{array}
 \eq
Straightforward calculation shows that it satisfies the quantum dynamical Yang-Baxter equation (\ref{w15}).
This $R$-matrix also satisfies the skew-symmetry (\ref{w06}) and unitarity (\ref{w05}) property.

In what follows we deal with the semi-dynamical ${\rm GL}_N$ rational $R$-matrix:
 \beq\label{w32}
 \begin{array}{c}
 \displaystyle{
 R^{\hbox{\tiny{Semi-dynam}}}_{12}(\hbar,z_1,z_2|\,q)=
 \sum\limits_{i\neq j}^N \Big( \frac{1}{z_1-z_2}+\frac{N}{q_j-q_i} \Big)E_{ij}\otimes E_{ji}
+ \sum\limits_{i\neq j}^N \Big( \frac{1}{\hbar}+\frac{N}{q_j-q_i} \Big)E_{ii}\otimes E_{jj}-
}
\\
 \displaystyle{
 -\sum\limits_{i\neq j}^N \Big( \frac{1}{z_1+\hbar}+\frac{N}{q_j-q_i} \Big)E_{ij}\otimes E_{jj}
 +\sum\limits_{i\neq j}^N \Big( \frac{1}{z_2}+\frac{N}{q_j-q_i} \Big)E_{jj}\otimes E_{ij}+
 }
\\
 \displaystyle{
 +\Big( \frac{1}{z_1-z_2}+\frac{1}{z_2}+\frac{1}{\hbar}-\frac{1}{z_1+\hbar} \Big)
 \sum\limits_{i=1}^N E_{ii}\otimes E_{ii}\,.
 }
 \end{array}
 \eq
It is also straightforward calculation to verify that it satisfies the quantum semi-dynamical Yang-Baxter equation
(\ref{w18}).
The skew-symmetry and unitarity properties are as follows:
 \beq\label{w33}
 \begin{array}{c}
 \displaystyle{
 R^{\hbox{\tiny{Semi-dynam}}}_{12}(\hbar,z_1,z_2|\,q)=
 -R^{\hbox{\tiny{Semi-dynam}}}_{21}(-\hbar,z_2+\hbar,z_1+\hbar|\,q)\,,
 }
 \end{array}
 \eq
 \beq\label{w34}
 \begin{array}{c}
 \displaystyle{
 R^{\hbox{\tiny{Semi-dynam}}}_{12}(\hbar,z_1,z_2|\,q)
 R^{\hbox{\tiny{Semi-dynam}}}_{21}(\hbar,z_2,z_1|\,q)=f(\hbar,z_1-z_2)1_N\otimes 1_N\,.
 }
 \end{array}
 \eq

Next, let us discuss relation between dynamical and semi-dynamical $R$-matrices (\ref{w31}) and (\ref{w32}).

\begin{theor}\label{theor1}
$R$-matrices (\ref{w31}) and (\ref{w32}) are related by the following twist transformation:
 \beq\label{w35}
 \begin{array}{c}
 \displaystyle{
 R^{\hbox{\tiny{Semi-dynam}}}_{12}(\hbar,z_1,z_2|\,q)=
 F_{12}(\hbar,z_1|\,q)\,
 R^{\hbox{\tiny{Dynam}}}_{12}(\hbar,z_1-z_2|\,q)\,
 F^{-1}_{21}(\hbar,z_2|\,q)\,,
 }
 \end{array}
 \eq
where
 \beq\label{w36}
 \begin{array}{c}
 \displaystyle{
 F_{12}(\hbar,z_1|\,q)=
  }
\\ \ \\
 \displaystyle{
 =\hbar\sum\limits_{i,j=1}^N\Big(\frac{1}{\hbar}-\frac{N}{q_i-q_j+N\hbar}\Big)E_{ii}\otimes E_{jj}
 +\hbar\sum\limits_{i,j=1}^N\Big(\frac{N}{q_i-q_j+N\hbar}-\frac{1}{z+\hbar}\Big)E_{ij}\otimes E_{jj}\,,
 }
 \end{array}
 \eq
 \beq\label{w37}
 \begin{array}{c}
 \displaystyle{
 F_{12}^{-1}(\hbar,z_1|\,q)=
 }
\\ \ \\
 \displaystyle{
 =\hbar\sum\limits_{i\neq j}^N\Big(\frac{1}{z}-\frac{N}{q_i-q_j}\Big)E_{ij}\otimes E_{jj}
 +\hbar\sum\limits_{i\neq j}^N\Big(\frac{N}{q_i-q_j}+\frac{1}{\hbar}\Big)E_{ii}\otimes E_{jj}
 +\hbar\Big(\frac{1}{\hbar}+\frac{1}{z}\Big)\sum\limits_{i=1}^N E_{ii}\otimes E_{ii}
 }
 \end{array}
 \eq
and
 \beq\label{w38}
 \begin{array}{c}
 \displaystyle{
 F_{12}(\hbar,z_1|\,q)=g_1^{-1}(z_1+\hbar,q)\,g_1(z_1,q-\hbar^{(2)})\,,
  }
 \end{array}
 \eq
so that the relation (\ref{w191}) holds true.
\end{theor}
The first part of the statement (\ref{w35})-(\ref{w37}) is a rational version of similar but more general elliptic relation from
\cite{Arut}. The second part of the statement (\ref{w38}) is deduced from
(\ref{e1674})-(\ref{e1675}).

\subsection{Vertex $R$-matrix from IRF-Vertex transformation}

\paragraph{Properties of gauged transformed semi-dynamical $R$-matrix.}

We are going to compute the vertex $R$-matrix using the semi-dynamical $R$-matrix (\ref{w32}) and the r.h.s. of
IRF-Vertex relation (\ref{w19}). First of all we need to prove that the expression
 \beq\label{w39}
  \begin{array}{c}
  \displaystyle{
 g_2(z_2,q)\,
 g_1(z_1+\hbar,q)\,
 R^{\hbox{\tiny{Semi-dynam}}}_{12}(\hbar,z_1,z_2|\,q)\, g_2^{-1}(z_2+\hbar,q)
 g_1^{-1}(z_1,q)\,.
 }
 \end{array}
 \eq
with the matrix $g(z,q)$ (\ref{e161}) is independent of dynamical parameters $q_1,...,q_N$.

\begin{predl}\label{prop1}
Expression (\ref{w39}) with the semi-dynamical $R$-matrix (\ref{w32}) and the twist matrix (\ref{e161})
is independent of variables $q_1,...,q_N$.
\end{predl}

\noindent\underline{\em Proof.} Fix some index $n$: $1\leq n\leq N$. Introduce the matrix
 \beq\label{w40}
  \begin{array}{c}
  \displaystyle{
 l^{(n)}(z,q)=g^{-1}(z,q)\p_{q_n} g(z,q)\in\Mat
 }
 \end{array}
 \eq
and
 \beq\label{w41}
  \begin{array}{c}
  \displaystyle{
 l(z,q)=g^{-1}(z,q)\p_{z} g(z,q)\in\Mat\,.
 }
 \end{array}
 \eq
The latter matrix is easily computed from (\ref{e1674})-(\ref{e1675}) since
 \beq\label{w42}
  \begin{array}{c}
  \displaystyle{
 \mL^\eta(z,q)=1_N-\eta\,l(z,q)+O(\eta^2)\,.
 }
 \end{array}
 \eq
This yields
 \beq\label{w43}
  \begin{array}{c}
  \displaystyle{
 l_{ij}(z,q)=\delta_{ij}\Big(\frac{1}{Nz}+\sum\limits_{k:k\neq j}^N\frac{1}{q_i-q_k}\Big)
 +(1-\delta_{ij})\Big(\frac{1}{Nz}-\frac{1}{q_i-q_j}\Big)\,.
 }
 \end{array}
 \eq
 Taking into account the form of dependence of matrix $g(z,q)$ on its variables (\ref{e161})-(\ref{e162})
  we also get
 explicit expression for $l^{(n)}(z,q)$:
 \beq\label{w44}
  \begin{array}{c}
  \displaystyle{
 l^{(n)}_{ij}(z,q)=\frac{N-1}{N}\,\delta_{jn}\,l_{ij}(z,q)
 -\frac{1}{N}\,(1-\delta_{jn})\,l_{ij}(z,q)-\delta_{ij}\,\frac{\p_{q_n}D_{ii}(q)}{D_{ii}(q)}\,,
 }
 \\
   \displaystyle{
 \frac{\p_{q_n}D_{ii}(q)}{D_{ii}(q)}=\frac{1-\delta_{in}}{q_n-q_i}+
 \delta_{in}\sum\limits_{k:k\neq i}^N\frac{1}{q_i-q_k}\,.
 }
 \end{array}
 \eq
 Let us compute the derivative of the expression (\ref{w39}) with respect to $q_n$:
 \beq\label{w45}
  \begin{array}{c}
  \displaystyle{
 g_2(z_2,q)\,
 g_1(z_1+\hbar,q)\Big(
 \p_{q_n}R^{\hbox{\tiny{Semi-dynam}}}_{12}(\hbar,z_1,z_2|\,q)
 +l^{(n)}_{1}(z_1+\hbar,q)\,R^{\hbox{\tiny{Semi-dynam}}}_{12}(\hbar,z_1,z_2|\,q)+
  }
 \\ \ \\
   \displaystyle{
 +l^{(n)}_{2}(z_2,q)\,R^{\hbox{\tiny{Semi-dynam}}}_{12}(\hbar,z_1,z_2|\,q)-
 R^{\hbox{\tiny{Semi-dynam}}}_{12}(\hbar,z_1,z_2|\,q)\,l^{(n)}_{2}(z_2+\hbar,q)-
 }
 \\ \ \\
   \displaystyle{
  -R^{\hbox{\tiny{Semi-dynam}}}_{12}(\hbar,z_1,z_2|\,q)\,l^{(n)}_{1}(z_1,q)
  \Big) g_2^{-1}(z_2+\hbar,q)\,g_1^{-1}(z_1,q)\,.
 }
 \end{array}
 \eq
Direct calculation shows that the expression inside the brackets in (\ref{w45}) is equal to zero. $\blacksquare$

\begin{predl}\label{prop2}
Expression (\ref{w39}) with the semi-dynamical $R$-matrix (\ref{w32}) and the twist matrix (\ref{e161})
depends o the difference of spectral parameters $z_1-z_2$ only.
\end{predl}

\noindent\underline{\em Proof.} This statement is proved similarly to the previous one. Compute the derivative
$\p_{z_1}+\p_{z_2}$ of the expression
(\ref{w39}):
$$
  \displaystyle{
 g_2(z_2,q)\,
 g_1(z_1+\hbar,q)\Big(
 (\p_{z_1}+\p_{z_2})R^{\hbox{\tiny{Semi-dynam}}}_{12}(\hbar,z_1,z_2|\,q)
 +l_{1}(z_1+\hbar,q)\,R^{\hbox{\tiny{Semi-dynam}}}_{12}(\hbar,z_1,z_2|\,q)+
  }
  $$
 \beq\label{w46}
  \begin{array}{c}
   \displaystyle{
 +l_{2}(z_2,q)\,R^{\hbox{\tiny{Semi-dynam}}}_{12}(\hbar,z_1,z_2|\,q)-
 R^{\hbox{\tiny{Semi-dynam}}}_{12}(\hbar,z_1,z_2|\,q)\,l_{2}(z_2+\hbar,q)-
 }
 \\ \ \\
   \displaystyle{
  -R^{\hbox{\tiny{Semi-dynam}}}_{12}(\hbar,z_1,z_2|\,q)\,l_{1}(z_1,q)
  \Big) g_2^{-1}(z_2+\hbar,q)\,g_1^{-1}(z_1,q)\,,
 }
 \end{array}
 \eq
where $l(z,q)$ is the matrix (\ref{w41})-(\ref{w43}). It is verified straightforwardly that
the expression inside the brackets in (\ref{w46}) vanishes. $\blacksquare$

Notice also that the expression (\ref{w39}) is a quantum $R$-matrix of vertex type.

\begin{predl}\label{prop3}
Expression (\ref{w39}) satisfies the quantum
Yang-Baxter equation (\ref{w01}).
\end{predl}
\noindent\underline{\em Proof.}
Indeed, it follows from Proposition \ref{prop2} that the expression (\ref{w39})
is equivalently written as
 \beq\label{w47}
  \begin{array}{c}
  \displaystyle{
 g_2(z_2-\hbar,q)\,
 g_1(z_1,q)\,
 R^{\hbox{\tiny{Semi-dynam}}}_{12}(\hbar,z_1-\hbar,z_2-\hbar|\,q)\, g_2^{-1}(z_2,q)
 g_1^{-1}(z_1-\hbar,q)\,.
 }
 \end{array}
 \eq
Plugging $R_{ab}^\hbar(z_a-z_b)$ into (\ref{w01}) in the forms (\ref{w39}) or (\ref{w47}) and one gets
 \beq\label{w48}
  \begin{array}{c}
  \displaystyle{
  R_{12}^\hbar(z_1-z_2) R_{13}^\hbar(z_1-z_3) R_{23}^\hbar(z_2-z_3)
   =g_1(z_1+\hbar,q)\, g_2(z_2,q)\, g_3(z_3-\hbar,q)\times
   }
 \\ \ \\
   \displaystyle{
\times
 \underline{
 R^{\hbox{\tiny{Semi-dynam}}}_{12}(\hbar,z_1,z_2|\,q)
 R^{\hbox{\tiny{Semi-dynam}}}_{13}(\hbar,z_1-\hbar,z_3-\hbar|\,q)
 R^{\hbox{\tiny{Semi-dynam}}}_{23}(\hbar,z_2,z_3|\,q)}\times
    }
 \\ \ \\
   \displaystyle{
 \times g_1^{-1}(z_1-\hbar,q)\,g_2^{-1}(z_2,q)\,g_3^{-1}(z_3+\hbar,q)\,.
 }
 \end{array}
 \eq
The underlined expression is the l.h.s. of (\ref{w18}). Making also similar calculation for the r.h.s. of (\ref{w01}) and
using the semi-dynamical Yang-Baxter equation (\ref{w18}) we conclude that (\ref{w01}) holds true. $\blacksquare$

To summarize, we proved

\begin{theor}\label{theor2}
Expression (\ref{w39}) is a quantum $R$-matrix of vertex type depending on the difference of spectral parameters.
\end{theor}

\paragraph{Explicit form of vertex-type rational $R$-matrix.} Let us slightly rewrite the semi-dynamical $R$-matrix (\ref{w32}):
 \beq\label{w49}
 \begin{array}{c}
  \displaystyle{
  R^{\hbox{\tiny{Semi-dynam}}}_{12}(\hbar,z_1,z_2|\,q)=\sum\limits_{i,j=1}^N\Big(\frac{E_{ij}\otimes E_{ji}}{z_1-z_2}+\frac{E_{ii}\otimes E_{jj}}{\hbar}
  -\frac{E_{ij}\otimes E_{jj}}{z_1+\hbar}+\frac{E_{jj}\otimes E_{ij}}{z_2}\Big)+
  }
\\
  \displaystyle{
  +N\sum\limits_{i\neq j}^N
  \Big(\frac{E_{ij}\otimes E_{ji}}{q_j-q_i}+\frac{E_{ii}\otimes E_{jj}}{q_j-q_i}
  -\frac{E_{ij}\otimes E_{jj}}{q_j-q_i}+\frac{E_{jj}\otimes E_{ij}}{q_j-q_i}\Big)\,.
    }
 \end{array}
 \eq
Then the expression (\ref{w39}) takes the form:
 \beq\label{w50}
 \begin{array}{c}
  \displaystyle{
  R^{\hbar}_{12}(z_1-z_2)=\sum\limits_{a,b,c,d=1}^N
  R^\hbar_{ab,cd}(z_1-z_2) E_{ab}\otimes E_{cd}\,,
    }
 \end{array}
 \eq
where
 \beq\label{w51}
 \begin{array}{c}
  \displaystyle{
  R^\hbar_{ab,cd}(z_1-z_2)=
    }
\\ \ \\
  \displaystyle{
  =\sum\limits_{i,j=1}^N\Big(
  \frac{ g_{ai}(z_1+\hbar)g_{cj}(z_2)g^{-1}_{jb}(z_1)g^{-1}_{id}(z_2+\hbar) }{z_1-z_2}
  +\frac{ g_{ai}(z_1+\hbar)g_{cj}(z_2)g^{-1}_{ib}(z_1)g^{-1}_{jd}(z_2+\hbar) }{\hbar}
    }
\\
  \displaystyle{
   \qquad\qquad\ -\frac{ g_{ai}(z_1+\hbar)g_{cj}(z_2)g^{-1}_{jb}(z_1)g^{-1}_{jd}(z_2+\hbar) }{z_1+\hbar}
  +\frac{ g_{aj}(z_1+\hbar)g_{ci}(z_2)g^{-1}_{jb}(z_1)g^{-1}_{jd}(z_2+\hbar) }{z_2}\Big)+
  }
  \\
    \displaystyle{
  +N\sum\limits_{i\neq j}^N\Big(
  \frac{ g_{ai}(z_1+\hbar)g_{cj}(z_2)g^{-1}_{jb}(z_1)g^{-1}_{id}(z_2+\hbar) }{q_j-q_i}
  +\frac{ g_{ai}(z_1+\hbar)g_{cj}(z_2)g^{-1}_{ib}(z_1)g^{-1}_{jd}(z_2+\hbar) }{q_j-q_i}
    }
\\
  \displaystyle{
   \qquad\qquad\ -\frac{ g_{ai}(z_1+\hbar)g_{cj}(z_2)g^{-1}_{jb}(z_1)g^{-1}_{jd}(z_2+\hbar) }{q_j-q_i}
  +\frac{ g_{aj}(z_1+\hbar)g_{ci}(z_2)g^{-1}_{jb}(z_1)g^{-1}_{jd}(z_2+\hbar) }{q_j-q_i}\Big)\,.
  }
 \end{array}
 \eq
Expressions for the matrix $g(z)$ and its inverse are given by (\ref{e161})-(\ref{e162})
and (\ref{e16722}) respectively.

\paragraph{Example 1.}
Since $R^\hbar_{ab,cd}(z_1-z_2)$ is independent of $q_1,...,q_N$ we may fix these variables in any possible way.
For example, let us fix
   \beq\label{w52}
 \begin{array}{c}
 \displaystyle{
q_i=i\,,\qquad {\bar q}_i=i-\frac{N+1}{2}\,.
}
 \end{array}
 \eq
Also, $R^\hbar_{ab,cd}(z_1-z_2)$ depends on the difference $z_1-z_2$ only. We may fix this freedom in different ways.
For instance,
   \beq\label{w53}
 \begin{array}{c}
 \displaystyle{
z_1=\frac{z}{2}\,,\quad z_2=-\frac{z}{2}\,.
}
 \end{array}
 \eq
Introduce the following set of functions:
   \beq\label{w540}
 \begin{array}{l}
 \displaystyle{
s_{kj}(z)=g^{-1}_{kj}(z_1,q)\Big|_{z_1=\frac{z}{2};\,q_i=i}\,,\qquad
t^\hbar_{kj}(z)=g^{-1}_{kj}(z_2+\hbar,q)\Big|_{z_2=-\frac{z}{2};\,q_i=i}
}
 \end{array}
 \eq
 and the following set of numbers:
   \beq\label{w55}
 \begin{array}{l}
 \displaystyle{
d_i=D_{ii}\Big|_{q_i=i}=\prod\limits_{k:k\neq i}^N(i-k)=(-1)^{N-i}(i-1)!(N-i)!\,.
}
 \end{array}
 \eq
Due to (\ref{e16722}) we have:
   \beq\label{w54}
 \begin{array}{l}
 \displaystyle{
s_{kj}(z)=(-1)^{\varrho(j)}\Big(\frac{2\sigma_{\varrho(j)}({y})}{Nz}\,-
\stackrel{k}{\sigma}_{\varrho(j)}({y})\Big)\,,\qquad y_j=\frac{z}{2}+j-\frac{N+1}{2}\,,
}
\\ \ \\
 \displaystyle{
t^\hbar_{kj}(z)=(-1)^{\varrho(j)}\Big(\frac{2\sigma_{\varrho(j)}({v})}{N(-z+2\hbar)}\,-
\stackrel{k}{\sigma}_{\varrho(j)}({v})\Big)\,,\qquad v_j=-\frac{z}{2}+\hbar+j-\frac{N+1}{2}\,.
}
 \end{array}
 \eq
Then (\ref{w51}) takes the following form:
 \beq\label{w56}
 \begin{array}{c}
  \displaystyle{
  R^\hbar_{ab,cd}(z)=
    }
\\ \ \\
  \displaystyle{
  =\sum\limits_{i,j=1}^N\Big(
  \frac{ (z/2+\hbar+{\bar q}_i)^{\rho(a)}(-z/2+{\bar q}_j)^{\rho(c)} }{d_i d_j}
  \Big[\frac{ s_{jb}(z)t^\hbar_{id}(z) }{z}+\frac{s_{ib}(z)t^\hbar_{jd}(z)}{\hbar}
  -\frac{2s_{jb}(z)t^\hbar_{jd}(z)}{z+2\hbar}\Big]-
    }
\\
  \displaystyle{
  -\frac{ (z/2+\hbar+{\bar q}_j)^{\rho(a)}(-z/2+{\bar q}_i)^{\rho(c)} }{d_i d_j}
  \frac{2s_{jb}(z)t^\hbar_{jd}(z)}{z}\Big)+
    }
 \end{array}
 \eq
$$
    \displaystyle{
  +N\sum\limits_{i\neq j}^N\Big(
  \frac{ (z/2+\hbar+{\bar q}_i)^{\rho(a)}(-z/2+{\bar q}_j)^{\rho(c)} }{d_i d_j(j-i)}
  \Big[ s_{jb}(z)t^\hbar_{id}(z) +s_{ib}(z)t^\hbar_{jd}(z)
  -s_{jb}(z)t^\hbar_{jd}(z)\Big]+
  }
$$
$$
    \displaystyle{
  +\frac{ (z/2+\hbar+{\bar q}_j)^{\rho(a)}(-z/2+{\bar q}_i)^{\rho(c)} }{d_i d_j}
  \frac{s_{jb}(z)t^\hbar_{jd}(z)}{j-i}\Big)\,.
  }
$$

\paragraph{Example 2.} Of course, the variables $z_1,z_2$ (and $q_1,...,q_N$) can be fixed in a different way.
Consider the following choice: $z_1=z$, $z_2=0$. Due to $R^{\hbox{\tiny{Semi-dynam}}}_{12}(\hbar,z_1,z_2|\,q)$
has a simple pole in $z_2$ at $z_2=0$ we should consider expression (\ref{w39}) carefully near $z_2=0$. The
semi-dynamical $R$-matrix has the following expansion:
 \beq\label{w561}
  \begin{array}{c}
  \displaystyle{
 R^{\hbox{\tiny{Semi-dynam}}}_{12}(\hbar,z_1,z_2|\,q)=\frac{1}{z_2}\,{\mathcal O}_{12}+{\mathcal B}_{12}(\hbar,z_1|q)
 +O(z_2)\,,
 }
 \end{array}
 \eq
 where
 \beq\label{w562}
  \begin{array}{c}
  \displaystyle{
 {\mathcal O}_{12}=\res\limits_{z_2=0}R^{\hbox{\tiny{Semi-dynam}}}_{12}(\hbar,z_1,z_2|\,q)=
 \sum\limits_{i,j=1}^N E_{jj}\otimes E_{ij}\,.
 }
 \end{array}
 \eq
 and 
 \beq\label{w563}
  \begin{array}{c}
  \displaystyle{
{\mathcal B}_{12}(\hbar,z_1|q)=\sum\limits_{i,j=1}^N\Big(\frac{E_{ij}\otimes E_{ji}}{z_1}+\frac{E_{ii}\otimes E_{jj}}{\hbar}
  -\frac{E_{ij}\otimes E_{jj}}{z_1+\hbar}\Big)+
  }
\\
  \displaystyle{
  +N\sum\limits_{i\neq j}^N
  \Big(\frac{E_{ij}\otimes E_{ji}}{q_j-q_i}+\frac{E_{ii}\otimes E_{jj}}{q_j-q_i}
  -\frac{E_{ij}\otimes E_{jj}}{q_j-q_i}+\frac{E_{jj}\otimes E_{ij}}{q_j-q_i}\Big)\,.
    }
 \end{array}
 \eq
Also
 \beq\label{w564}
  \begin{array}{c}
  \displaystyle{
  g_2(z_2,q)=g_2(0,q)+z_2g'_2(0,q)+O(z_2^2)\,,\qquad g'_2(z,q)=\p_z g_2(z,q)\,.
 }
 \end{array}
 \eq
 Plugging all expansions into (\ref{w39}) we get
 \beq\label{w565}
  \begin{array}{c}
  \displaystyle{
 R_{12}^\hbar(z)=
 g_1(z+\hbar,q)\Big(g'_2(0,q){\mathcal O}_{12}+g_2(0,q){\mathcal B}_{12}(\hbar,z|q)\Big)
  g_2^{-1}(\hbar,q)
 g_1^{-1}(z,q)\,,
 }
 \end{array}
 \eq
 where we used the following relation 
 \beq\label{w566}
  \begin{array}{c}
  \displaystyle{
g_2(0){\mathcal O}_{12}=0\,,
 }
 \end{array}
 \eq
 which is valid due to the property (\ref{e16733}).
 
Similar representation exists for trigonometric and elliptic $R$-matrices.

\subsection{Coincidence of old and new forms of rational $R$-matrix}

Here we explain why expression (\ref{w51}) coincides with the one (\ref{w80}).
 A direct proof is
complicated. For finite $N$ one can use computer calculations to verify the coincidence 
(we made it for $N=2,...,6$). For an arbitrary $N$ we will prove the coincidence using the fact that
(\ref{w80}) was derived in \cite{LOZ14} through the classical analogue of the
IRF-Vertex relation (gauge equivalence of classical Lax matrices). Relation between the classical and quantum IRF-Vertex relations was clarified in \cite{VZ} for elliptic models. Below we use similar approach.

In order to distinguish (\ref{w51}) and (\ref{w80}) 
we denote the expression (\ref{w80}) as ${\ti R}^\hbar_{12}(z)$.
Let us briefly recall idea of derivation of (\ref{w80}). 
The factorization property (\ref{e1675}) allows to write the Lax matrix of the classical Ruijsenaars-Schneider model
as
 \beq\label{w5670}
  \begin{array}{c}
  \displaystyle{
L^{\hbox{\tiny{RS}}}(z)=g^{-1}(z,q)g(z+\eta,q)e^P\,,
 }
 \end{array}
 \eq
 where $P={\rm diag}(p_1,...,p_N)\in\Mat$ is a diagonal matrix of momenta. The variables $q_1,...,q_N$ play the role of positions of particles. Perform the gauge transformation
 \beq\label{w5671}
  \begin{array}{c}
  \displaystyle{
L^{\hbox{\tiny{top}}}(z)=g(z,q)L^{\hbox{\tiny{RS}}}(z)g^{-1}(z,q)=g(z+\eta,q)e^P g^{-1}(z,q)
 }
 \end{array}
 \eq
 and compute the residue of the obtained expression:
 \beq\label{w5672}
  \begin{array}{c}
  \displaystyle{
S=S(p,q,\eta)=\res\limits_{z=0}L^{\hbox{\tiny{top}}}(z)=g(\eta,q)e^P {\breve g}(0,q)\in\Mat\,,\qquad 
{\breve g}(0,q)=\res\limits_{z=0} g^{-1}(z)\,.
 }
 \end{array}
 \eq
Main observation is that $L^{\hbox{\tiny{top}}}(z)$ (\ref{w5671}) is represented in the following 
form\footnote{The Lax matrix of the form (\ref{w5672}) describes the so-called relativistic integrable top model \cite{LOZ14}.}:
 \beq\label{w5673}
  \begin{array}{c}
  \displaystyle{
L^{\hbox{\tiny{top}}}(z)=\tr_2\Big({\ti R}^\eta_{12}(z) S_2\Big)\,,\qquad S_2=1_N\otimes S\,,
 }
 \end{array}
 \eq
where $\tr_2$ is a trace over the second tensor component and ${\ti R}^\eta_{12}(z)$ is independent of the variables $p_1,,,.p_N$ and $q_1,...,q_N$. The expression (\ref{w80}) was computed in this way, i.e. relation (\ref{w5673})
can be viewed as definition of ${\ti R}^\eta_{12}(z)$.

Let us prove that the relation (\ref{w5673}) holds true for the $R$-matrix $R^\eta_{12}(z)$ (\ref{w51}).
For this purpose we need one more property of (\ref{w51}) mentioned in \cite{SeZ}. Recall that $R^\eta_{12}(z)$ was derived as given in (\ref{w19}).
Multiply both sides of (\ref{w19}) by the matrix $g_2^{-1}(z_2)$ from the left
 \beq\label{w5674}
  \begin{array}{c}
  \displaystyle{
 g^{-1}_2(z_2,q)\, R^\hbar_{12}(z_1-z_2)=
 g_1(z_1+\hbar,q)\,
 R^{\hbox{\tiny{Semi-dynam}}}_{12}(\hbar,z_1,z_2|\,q)\, g_2^{-1}(z_2+\hbar,q)
 g_1^{-1}(z_1,q)
 }
 \end{array}
 \eq
and calculate the residue of both sides in $z_2$ variable at $z_2=0$ using (\ref{w562}) and notation ${\breve g}(0,q)$ 
from (\ref{w5672}):
 \beq\label{w5675}
  \begin{array}{c}
  \displaystyle{
 {\breve g}_2(0,q)\, R^\hbar_{12}(z)=
 g_1(z+\hbar,q)\,
 {\mathcal O}_{12}\, g_2^{-1}(\hbar,q)
 g_1^{-1}(z,q)\,.
 }
 \end{array}
 \eq

\begin{predl}\label{prop34}
 The rational vertex type ${\rm GL}_N$ $R$-matrix (\ref{w51}) satisfies
 relation (\ref{w5673}).
\end{predl}

\noindent \underline{\em Proof.} Notice that
 \beq\label{w5676}
  \begin{array}{c}
  \displaystyle{
 \tr_2\Big( {\mathcal O}_{12}\,e^{P_2} \Big)=e^P\,.
 }
 \end{array}
 \eq
Therefore, 
 \beq\label{w5677}
  \begin{array}{c}
  \displaystyle{
 g(z+\eta,q)e^P g^{-1}(z,q)=\tr_2\Big( g_1(z+\eta,q){\mathcal O}_{12}\,e^{P_2}g_1^{-1}(z,q) \Big)=
 }
 \\ \ \\
   \displaystyle{
=\tr_2\Big( g_1(z+\eta,q){\mathcal O}_{12}g_1^{-1}(z,q)g_2^{-1}(\eta,q) g_2(\eta,q)e^{P_2} \Big)\stackrel{(\ref{w5675})}{=}
\tr_2\Big( {\breve g}_2(0,q)\, R^\eta_{12}(z) g_2(\eta,q)e^{P_2} \Big)=
 }
  \\ \ \\
   \displaystyle{
   =\tr_2\Big(  R^\eta_{12}(z) S_2  \Big)\,.
   }
 \end{array}
 \eq
In this way we showed that two definitions (\ref{w51}) and (\ref{w80}) of the rational $R$-matrix coincide.
$\blacksquare$

\section{Associative Yang-Baxter equation and other $R$-matrix properties}
\setcounter{equation}{0}

\subsection{Skew-symmetry and unitarity}

\begin{predl}\label{prop41}
 The rational vertex type ${\rm GL}_N$ $R$-matrix (\ref{w51}) or (\ref{w80}) satisfies
the unitarity (\ref{w05}) and the skew-symmetry (\ref{w06}) properties.
\end{predl}

\noindent \underline{\em Proof.} Plugging (\ref{w39}) into $R_{12}^\hbar(z)R_{21}^\hbar(-z)$ we get
   \beq\label{w57}
 \begin{array}{c}
 \displaystyle{
R_{12}^\hbar(z_1-z_2)R_{21}^\hbar(z_2-z_1)=
  }
  \\ \ \\
    \displaystyle{
=g_2(z_2,q)\,
 g_1(z_1+\hbar,q)\,
 R^{\hbox{\tiny{Semi-dynam}}}_{12}(\hbar,z_1,z_2|\,q)\, g_2^{-1}(z_2+\hbar,q)
 g_1^{-1}(z_1,q)\times
   }
  \\ \ \\
    \displaystyle{
   \times g_1(z_1,q)\,
 g_2(z_2+\hbar,q)\,
 R^{\hbox{\tiny{Semi-dynam}}}_{21}(\hbar,z_2,z_1|\,q)\, g_1^{-1}(z_1+\hbar,q)
 g_2^{-1}(z_2,q)\stackrel{(\ref{w34})}{=}
}
  \\ \ \\
    \displaystyle{
=f(\hbar,z_1-z_2)1_N\otimes 1_N\,.
}
 \end{array}
 \eq
 Let us write down the semi-dynamical $R$-matrix in terms of the vertex one:
   \beq\label{w58}
 \begin{array}{l}
 \displaystyle{
R^{\hbox{\tiny{Semi-dynam}}}_{12}(\hbar,z_1,z_2|\,q)=g_2^{-1}(z_2,q)\,
 g_1^{-1}(z_1+\hbar,q)\,
 R_{12}^\hbar(z_1-z_2)\, g_2(z_2+\hbar,q)
 g_1(z_1,q)\,.
}
 \end{array}
 \eq
Then
   \beq\label{w59}
 \begin{array}{l}
 \displaystyle{
 R^{\hbox{\tiny{Semi-dynam}}}_{21}(-\hbar,z_2+\hbar,z_1+\hbar|\,q)=g_1^{-1}(z_1+\hbar,q)\,
 g_2^{-1}(z_2,q)\,
 R_{21}^{-\hbar}(z_2-z_1)
 g_1(z_1,q)\, g_2(z_2+\hbar,q)
}
 \end{array}
 \eq
and by comparing (\ref{w58}) and (\ref{w59}) we get (\ref{w06}) due to (\ref{w33}). $\blacksquare$

\begin{predl}\label{prop42}
The rational vertex type ${\rm GL}_N$ $R$-matrix (\ref{w51}) or (\ref{w80}) has the following local behaviour:
   \beq\label{w60}
 \begin{array}{l}
 \displaystyle{
 \res\limits_{\hbar=0}R_{12}^\hbar(z)=1_N\otimes 1_N\,,\qquad
 \res\limits_{z=0}R_{12}^\hbar(z)=P_{12}\,.
}
 \end{array}
 \eq
\end{predl}
The proof easily follows from representing $R_{12}^\hbar(z)$ in the form (\ref{w39}) and
(also easily obtained) the following properties of the semi-dynamical $R$-matrix (these properties are easily verified):
   \beq\label{w61}
 \begin{array}{l}
 \displaystyle{
 \res\limits_{\hbar=0}R^{\hbox{\tiny{Semi-dynam}}}_{12}(\hbar,z_1,z_2|\,q)=1_N\otimes 1_N\,,\qquad
 \res\limits_{z_1=z_2}R^{\hbox{\tiny{Semi-dynam}}}_{12}(\hbar,z_1,z_2|\,q)=P_{12}\,.
}
 \end{array}
 \eq
 The same result follows from the second line of (\ref{w51}). Notice that the vertex $R$-matrix has no
 higher order poles in $\hbar$. Therefore, we proved that this $R$-matrix obeys
 the classical limit expansion (\ref{w08}) near $\hbar=0$.
 
\subsection{Symmetry of arguments} 

The elliptic Baxter-Belavin $R$-matrix (in the fundamental representation of ${\rm GL}_N$ Lie group) satisfies also the symmetry of arguments property:
   \beq\label{w62}
 \begin{array}{l}
 \displaystyle{
 R_{12}^\hbar(z)P_{12}=R_{12}^{\, z}(\hbar)\,.
}
 \end{array}
 \eq
 It is very useful in different calculations. Let us prove it for the rational case (\ref{w51}). As in all upper statements, our strategy
 is to find analogue of (\ref{w62}) for the semi-dynamical $R$-matrix (\ref{w49}) and then use the IRF-Vertex relation (\ref{w58}).
 The semi-dynamical analogue of (\ref{w62}) is as follows.
 \begin{lemma}\label{lem41}
 The semi-dynamical $R$-matrix (\ref{w49}) satisfies the following analogue
 of the symmetry of arguments property\footnote{We could not find this property in the literature.}:
   \beq\label{w63}
 \begin{array}{l}
 \displaystyle{
 R^{\hbox{\tiny{Semi-dynam}}}_{12}(\hbar,z_1,z_2|\,q)
 =R^{\hbox{\tiny{Semi-dynam}}}_{12}(z_1-z_2,\hbar+z_2,z_2|\,q)P_{12}\,.
}
 \end{array}
 \eq
 \end{lemma}
  \noindent \underline{\em Proof.} It is verified straightforwardly. Using the properties of the action 
  of permutation operator
   \beq\label{w64}
 \begin{array}{c}
 \displaystyle{
E_{ij}\otimes E_{ji}P_{12}=E_{ii}\otimes E_{jj}\,,\qquad E_{ii}\otimes E_{jj}P_{12}=E_{ij}\otimes E_{ji}\,,
}
\\ \ \\
 \displaystyle{
E_{ij}\otimes E_{jj}P_{12}=E_{ij}\otimes E_{jj}\,,\qquad E_{jj}\otimes E_{ij}P_{12}=E_{jj}\otimes E_{ij}\,,
\qquad E_{ii}\otimes E_{ii}P_{12}=E_{ii}\otimes E_{ii}
}
 \end{array}
 \eq
 one gets (\ref{w63}). $\blacksquare$

 \begin{predl}\label{prop43}
 The rational vertex type ${\rm GL}_N$ $R$-matrix (\ref{w51}) or (\ref{w80}) satisfies the symmetry of arguments property (\ref{w62}).
 \end{predl}
 \noindent \underline{\em Proof.} Due to (\ref{w58}) for the r.h.s. of (\ref{w63}) we have
   \beq\label{w65}
 \begin{array}{l}
  \displaystyle{
R^{\hbox{\tiny{Semi-dynam}}}_{12}(z_1-z_2,\hbar+z_2,z_2|\,q)P_{12}=
}
 \\ \ \\
 \displaystyle{
=g_2^{-1}(z_2,q)\,
 g_1^{-1}(z_1+\hbar,q)\,
 R_{12}^{z_1-z_2}(\hbar)\, g_2(z_1,q)
 g_1(z_2+\hbar,q)\,P_{12}=
}
\\ \ \\
 \displaystyle{
=g_2^{-1}(z_2,q)\,
 g_1^{-1}(z_1+\hbar,q)\,
 R_{12}^{z_1-z_2}(\hbar)\,P_{12}\, g_1(z_1,q)
 g_2(z_2+\hbar,q)\,.
}
 \end{array}
 \eq
 The l.h.s. of (\ref{w63}) is given by (\ref{w58}). By comparing (\ref{w58}) and (\ref{w65}) the statement is proved.
 $\blacksquare$


\subsection{Associative and quantum Yang-Baxter equations}
The proof of the associative Yang-Baxter equation (\ref{w14}) for the vertex $R$-matrix in the elliptic case is quite simple (see e.g. \cite{LOZ141}). In the rational case a direct proof is a technically complicated task.
For this reason we use the associative Yang-Baxter equation (\ref{w20}) for semi-dynamical  $R$-matrix and perform the 
gauge (IRF-Vertex) transformation. In fact, (\ref{w20}) was originally derived from (\ref{w14}) in the same way \cite{SeZ}.

 \begin{predl}\label{prop44}
 The rational vertex type ${\rm GL}_N$ $R$-matrix (\ref{w51}) or (\ref{w80}) satisfies the associative Yang-Baxter equation (\ref{w14}).
 \end{predl}
  \noindent \underline{\em Proof.}
Plugging (\ref{w58}) into the terms from semi-dynamical associative Yang-Baxter equation (\ref{w20}) we get:
  \beq\label{w66}
  \begin{array}{c}
  \displaystyle{
 R^{\hbox{\tiny{Semi-dynam}}}_{12}(\hbar ,z_1+\eta,z_2+\eta)
 R^{\hbox{\tiny{Semi-dynam}}}_{23}({\eta},z_2+\hbar,z_3+\hbar)=
 }
 \end{array}
 \eq
 $$
   \displaystyle{
   g_1^{-1}(z_1+\hbar+\eta)g_2^{-1}(z_2+\eta)g_3^{-1}(z_3+\hbar)
 R^{\hbar}_{12}(z_1-z_2) R^{\eta}_{23}(z_2-z_3)
 g_1(z_1+\eta)g_2(z_2+\hbar)g_3(z_3+\hbar+\eta)\,,
 }
 $$
  \beq\label{w67}
  \begin{array}{c}
  \displaystyle{
 R^{\hbox{\tiny{Semi-dynam}}}_{13}({\eta},z_1+\hbar,z_3+\hbar)
 R^{\hbox{\tiny{Semi-dynam}}}_{12}({\hbar-\eta},z_1+\eta,z_2+\eta)=
 }
  \end{array}
 \eq
 $$
   \displaystyle{
   g_1^{-1}(z_1+\hbar+\eta)g_2^{-1}(z_2+\eta)g_3^{-1}(z_3+\hbar)
 R^{\eta}_{13}(z_1-z_3) R^{\hbar-\eta}_{12}(z_1-z_2)
  g_1(z_1+\eta)g_2(z_2+\hbar)g_3(z_3+\hbar+\eta)\,,
 }
 $$
  \beq\label{w68}
  \begin{array}{c}
  \displaystyle{
 R^{\hbox{\tiny{Semi-dynam}}}_{23}({\eta-\hbar},z_2+\hbar,z_3+\hbar)
 R^{\hbox{\tiny{Semi-dynam}}}_{13}(\hbar,z_1+\eta,z_3+\eta)=
 }
  \end{array}
 \eq
 $$
   \displaystyle{
   g_1^{-1}(z_1+\hbar+\eta)g_2^{-1}(z_2+\eta)g_3^{-1}(z_3+\hbar)
 R^{\eta-\hbar}_{23}(z_2-z_3) R^{\hbar}_{13}(z_1-z_3)
  g_1(z_1+\eta)g_2(z_2+\hbar)g_3(z_3+\hbar+\eta)\,.
 }
 $$
Then (\ref{w14}) follows from (\ref{w20}). $\blacksquare$

Together with the properties (\ref{w05})-(\ref{w08}) the associative Yang-Baxter equations provides 
a wide set of identities, which are useful in different applications. For example, the classical 
$r$-matrix satisfies not only the classical Yang-Baxter equation but also the following relation:
  \beq\label{w69}
  \begin{array}{c}
  \displaystyle{
 \Big(r_{12}(z_1-z_2)+r_{23}(z_2-z_3)+r_{31}(z_3-z_1)\Big)^2=
 }
 \\ \ \\
   \displaystyle{
 =1_N\otimes 1_N\otimes 1_N
 \Big( \frac{1}{(z_1-z_2)^2}+\frac{1}{(z_2-z_3)^2}+\frac{1}{(z_3-z_1)^2} \Big)\,.
 }
  \end{array}
 \eq
The coefficient $m_{12}(z)$ in the expansion (\ref{w08}) can be computed as
  \beq\label{w70}
  \begin{array}{c}
  \displaystyle{
 m_{12}(z)=\frac12\Big(r^2_{12}(z)-\frac{1_N\otimes 1_N}{z^2}\Big)\,.
 }
  \end{array}
 \eq
Many other identities can be found in \cite{AtZ,LOZ141,LOZ16,MZ,TrZ}.

Finally, the quantum Yang-Baxter equation (\ref{w01}) for the rational ${\rm GL}_N$ vertex type $R$-matrix is by construction, see Proposition \ref{prop3}. Alternatively, one can derive (\ref{w01}) from the associative Yang-Baxter
equation (\ref{w14}) and the properties (\ref{w05})-(\ref{w06}). A simple proof can be found in \cite{LOZ141}.

\section{Appendix: Rational $R$-matrix in ${\rm GL}_N$ case}\label{app}
\def\theequation{A.\arabic{equation}}
\setcounter{equation}{0}

Here we write down the rational ${\rm GL}_N$ $R$-matrix from \cite{LOZ14} and \cite{AASZ}.

\subsection{Quantum $R$-matrix}

We begin with the quantum $R$-matrix. Below is slightly modified and corrected version of the one obtained in
\cite{LOZ14}. All the sums below should be primed (we do not put primes just to keep the place) in the sense that
the summation indices run over all values, which are well defined for the corresponding expressions. More precisely,
the summands contain expressions of the form $\rho^{-1}(m)$. Due to (\ref{e163}) such expressions are defined for
$m=1,...,N-2$ and $m=N$ but are not defined for $m=N-1$. The corresponding terms are skipped in the sums below.
The answer for $R$-matrix is as follows:
 \beq\label{w80}
 \begin{array}{c}
 \displaystyle{
R_{12}^{\hbar}(z)=A(z,\hbar)\otimes B(\hbar)+\frac{1}{z} \sum_{i, j=1}^{N} {E}_{i j} \otimes\left\{\sum_{\gamma=0}^{\varrho(i)} z^{\gamma}\left(\begin{array}{c}\varrho(i) \\ \gamma\end{array}\right)
{E}_{j, \varrho^{-1}(\varrho(i)-\gamma)}-\right.
 }
 \\ \ \\
  \displaystyle{
  -\sum_{\gamma=0}^{\varrho(i)} z^{\gamma+N-j+1}(-1)^{\varrho(j)+N}(N-j)\left(\begin{array}{c}\varrho(i) \\ \gamma\end{array}\right)\left(\begin{array}{c}N \\ j-1\end{array}\right) {E}_{N, \varrho^{-1}(\varrho(i)-\gamma)}+
  }
   \end{array}
 \eq
   $$
  \displaystyle{
+\sum_{\gamma=0}^{\varrho(i)} \sum_{s=1}^{N-j} z^{s+\gamma}(-1)^{s+\delta_{j,N}}\left(\begin{array}{c}\varrho(i) \\ \gamma\end{array}\right)\left(\begin{array}{c}s+j-1 \\ j-1\end{array}\right) {E}_{\varrho^{-1}(s+j-1), \varrho^{-1}(\varrho(i)-\gamma)}-
  }
   $$
   $$
  \displaystyle{
-N \sum_{\gamma=0}^{\varrho(i)} \sum_{s=1}^{N-j}(-1)^{s+\delta_{j,N}} z^{s}(z+\hbar)^{\gamma}\left(\begin{array}{c}s+j-2 \\ j-1\end{array}\right)\left(\begin{array}{c}\varrho(i) \\ \gamma\end{array}\right) \times
  }
 $$
$$\times\left[\delta_{\varrho(i)+1\leq j+s+\gamma} \sum_{c=0}^{N-s-j+1} \sum_{p=0}^{\varrho(i)-\gamma+c}(-\hbar)^{p}\left(\begin{array}{c}\varrho(i)-\gamma+c \\ p\end{array}\right) {E}_{\varrho^{-1}(s+j+c-1), \varrho^{-1}(\varrho(i)-\gamma-p+c)}-\right.
$$
$$\left.\left.-\delta_{\varrho(i)+1>j+s+\gamma} \sum_{c=0}^{s+j-2} \sum_{p=0}^{\varrho(i)-\gamma-c-1}\!(-\hbar)^{p}\left(\!\begin{array}{c}\!\varrho(i)\!-\!\gamma\!-\!c\!-\!1\!\\ \!p\!\end{array}\!\right) {E}_{\varrho^{-1}(s+j-c-2), \varrho^{-1}(\varrho(i)-\gamma-p-c-1)}\right]\right\}
\,,
$$
where are $A(z,\hbar)$ and $B(\hbar)$ are the following ${\mathrm{Mat}}_N(\mathbb C)$-valued functions:
 \beq\label{w81}
 \begin{array}{c}
 \displaystyle{
A(z,\hbar)={E}_{N N}-\sum_{j=1}^{N} (N-j)z^{N-j+1}(-1)^{\varrho(j)+N} \left(\begin{array}{c}N \\ j-1\end{array}\right) {E}_{N j}-
   }
 \end{array}
 \eq
$$
- \sum_{i, j=1}^{N} \sum_{s=1}^{N-j} \sum_{b=0}^{\varrho(i)}(-1)^{s+\delta_{j,N}} z^{s-1}(z+\hbar)^{b}\left(\!\begin{array}{c}\!s\!+\!j\!-\!2\! \\ j-1\end{array}\!\right)\left(\!\begin{array}{c}\!\varrho(i)\! \\ \!b\!\end{array}\!\right)
\Big(\delta_{\varrho(i)-j,b+s-2}-N\hbar\delta_{\varrho(i)-j,b+s-1}\Big) {E}_{i j}\,,
$$

$$B(\hbar)=\frac{1_N}{\hbar}-\frac{1}{N} \sum_{j=1}^{N}\left[\delta_{\varrho(j) \geq 1} \varrho(j) {E}_{j, \varrho^{-1}(\varrho(j)-1)}-(-1)^{\delta_{j,N}} j {E}_{\varrho^{-1}(j), j}
+N\sum_{b=0}^{\varrho(j)}(-1)^{b+\delta_{j,N}}\times\right.$$
 \beq\label{w82}
 \begin{array}{c}
 \displaystyle{
\left. \times\left(\begin{array}{c}\varrho(j) \\ b\end{array}\right) \sum_{c=0}^{N-j} \sum_{p=0}^{\varrho(j)-b+c}(-\hbar)^{p+b}\left(\begin{array}{c}\varrho(j)-b+c \\  p\end{array}\right) {E}_{\varrho^{-1}(j+c), \varrho^{-1}(\varrho(j)-b-p+c)}\right]\,.
   }
 \end{array}
 \eq

\subsection{Classical $r$-matrix}

The classical $r$-matrix
 \beq\label{w83}
 \begin{array}{c}
 \displaystyle{
r_{12}(z)=\lim\limits _{\hbar \rightarrow 0}\left(R_{12}(z)-1_{N} \otimes 1_{N} / \hbar\right)
   }
 \end{array}
 \eq
 was computed in \cite{AASZ}. Here a slightly different expression obtained from (\ref{w80}). Consider expansions
 of matrices $A(z,\hbar)$ (\ref{w81}) and $B(\hbar)$ (\ref{w82}):
 \beq\label{w84}
 \begin{array}{c}
 \displaystyle{
A(z,\hbar)=A^{[0]}(z)+\hbar A^{[1]}(z)+\hbar^2 A^{[2]}(z)...
   }
   \\ \ \\
  \displaystyle{
    B(\hbar)=\hbar^{-1}1_N+B^{[0]}+\hbar B^{[1]}+...
    }
 \end{array}
 \eq
The answer for the classical $r$-matrix is as follows:
 \beq\label{w85}
 \begin{array}{c}
 \displaystyle{
r_{12}(z)=A^{[0]}(z)\otimes B^{[0]}+A^{[1]}(z)\otimes 1_N+\frac{1}{z} \sum_{i, j=1}^{N} {E}_{i j} \otimes\left\{\sum_{\gamma=0}^{\varrho(i)} z^{\gamma}\left(\begin{array}{c}\varrho(i) \\ \gamma\end{array}\right)
{E}_{j, \varrho^{-1}(\varrho(i)-\gamma)}-\right.
}
\\ \ \\
  \displaystyle{
-\sum_{\gamma=0}^{\varrho(i)} z^{\gamma+N-j+1}(-1)^{\varrho(j)+N}(N-j)\left(\begin{array}{c}\varrho(i) \\ \gamma\end{array}\right)\left(\begin{array}{c}N \\ j-1\end{array}\right) {E}_{N, \varrho^{-1}(\varrho(i)-\gamma)}+
}
 \end{array}
 \eq
$$
  \displaystyle{
+\sum_{\gamma=0}^{\varrho(i)} \sum_{s=1}^{N-j} z^{s+\gamma}(-1)^{s+\delta_{j,N}}\left(\begin{array}{c}\varrho(i) \\ \gamma\end{array}\right)\left(\begin{array}{c}s+j-1 \\ j-1\end{array}\right) {E}_{\varrho^{-1}(s+j-1), \varrho^{-1}(\varrho(i)-\gamma)}-
}
$$
$$
  \displaystyle{
-N \sum_{\gamma=0}^{\varrho(i)} \sum_{s=1}^{N-j}(-1)^{s+\delta_{j,N}} z^{s+\gamma}
\left(\begin{array}{c}s+j-2 \\ j-1\end{array}\right)\left(\begin{array}{c}\varrho(i) \\ \gamma\end{array}\right)
\Big[\delta_{\varrho(i)+1\leq j+s+\gamma}\times
    }
$$
$$ \times\sum_{c=0}^{N-s-j+1} {E}_{\varrho^{-1}(s+j+c-1), \varrho^{-1}(\varrho(i)-\gamma+c)}
\left.-\delta_{\varrho(i)+1>j+s+\gamma} \sum_{c=0}^{s+j-2} {E}_{\varrho^{-1}(s+j-c-2), \varrho^{-1}(\varrho(i)-\gamma-c-1)}\Big]\right\}
\,,
$$
where are $A^{[0]}(z)$, $A^{[1]}(z)$ and $B^{[0]}$ are the following ${\mathrm{Mat}}_N(\mathbb C)$-valued functions:
 \beq\label{w86}
 \begin{array}{c}
 \displaystyle{
A^{[0]}(z)={E}_{N N}-\sum_{j=1}^{N} (N-j)z^{N-j+1}(-1)^{\varrho(j)+N} \left(\begin{array}{c}N \\ j-1\end{array}\right) {E}_{N j}-
}
\\ \ \\
  \displaystyle{
- \sum_{i, j=1}^{N} \sum_{s=1}^{N-j} \sum_{b=0}^{\varrho(i)}(-1)^{s+\delta_{j,N}} z^{s+b-1}\left(\!\begin{array}{c}\!s\!+\!j\!-\!2\! \\ j-1\end{array}\!\right)\left(\!\begin{array}{c}\!\varrho(i)\! \\ \!b\!\end{array}\!\right)
\delta_{\varrho(i)-j,b+s-2} {E}_{i j}\,,
    }
 \end{array}
 \eq
 \beq\label{w87}
 \begin{array}{c}
 \displaystyle{
A^{[1]}(z)=- \sum_{i, j=1}^{N} \sum_{s=1}^{N-j} \sum_{b=0}^{\varrho(i)}(-1)^{s+\delta_{j,N}} z^{s+b-2}\left(\!\begin{array}{c}\!s\!+\!j\!-\!2\! \\ j-1\end{array}\!\right)\left(\!\begin{array}{c}\!\varrho(i)\! \\ \!b\!\end{array}\!\right)\times
}
\\ \ \\
  \displaystyle{
\times\Big(b\,\delta_{\varrho(i)-j,b+s-2}-Nz\,\delta_{\varrho(i)-j,b+s-1}\Big) {E}_{i j}\,,
    }
 \end{array}
 \eq
 \beq\label{w88}
 \begin{array}{c}
 \displaystyle{
B^{[0]}=-\frac{1}{N} \sum_{j=1}^{N}\Big[\delta_{\varrho(j) \geq 1} \varrho(j) {E}_{j, \varrho^{-1}(\varrho(j)-1)}-(-1)^{\delta_{j,N}} j {E}_{\varrho^{-1}(j), j}
+
}
\\ \ \\
  \displaystyle{
+N(-1)^{\delta_{j,N}}
\sum_{c=0}^{N-j}  {E}_{\varrho^{-1}(j+c), \varrho^{-1}(\varrho(j)+c)}\Big]\,.
    }
 \end{array}
 \eq

\subsection{Other coefficients}

For different applications (see ...) some other coefficients of $R$-matrix expansion are needed. Here we compute
some of them. Let us begin with $m_{12}(z)$ matrix entering (\ref{w08}).

\underline{Matrix $m_{12}(z)$:}

$$m_{12}(z)=A^{[0]}(z)\otimes B^{[1]}+A^{[1]}(z)\otimes B^{[0]}+A^{[2]}(z)\otimes 1_N+ \sum_{i, j=1}^{N} {E}_{i j} \otimes\left\{N \sum_{s=1}^{N-j} \sum_{\gamma =0}^{\varrho(i)}(-1)^{s+\delta_{j,N}}z^{s+\gamma-2}\times \right.$$
%
$$\times \left(\!\begin{array}{c}\!s\!+\!j\!-\!2\! \\ j-1\end{array}\!\right)\left(\!\begin{array}{c}\!\varrho(i)\! \\ \!\gamma \!\end{array}\!\right)\left[ \delta_{\varrho(i)+1\leq j+s+\gamma} \sum_{c=0}^{N-s-j+1} \Big(z (\varrho(i)-\gamma +c){E}_{\varrho^{-1}(s+j+c-1), \varrho^{-1}(\varrho(i)-\gamma+c-1)}- \right.$$
$$-\gamma {E}_{\varrho^{-1}(s+j+c-1), \varrho^{-1}(\varrho(i)-\gamma+c)} \Big)
 +\delta_{\varrho(i)+1> j+s+\gamma} \sum_{c=0}^{s+j-2} \Big(\gamma {E}_{\varrho^{-1}(s+j-c-2), \varrho^{-1}(\varrho(i)-\gamma-c-1)}-
 $$
 \beq\label{w89}
 \begin{array}{c}
 \displaystyle{
\left.\left. \phantom{\sum_{s=1}^{N-j}}- z (\varrho(i)-\gamma-c-1)  {E}_{\varrho^{-1}(s+j-c-2), \varrho^{-1}(\varrho(i)-\gamma-c-2)} \Big)\right]\right\}
\,,
    }
 \end{array}
 \eq
where $A^{[2]}(z)$  and $B^{[1]}$ are the following matrices:
 \beq\label{w90}
 \begin{array}{c}
 \displaystyle{
A^{[2]}(z)= -\sum_{i, j=1}^{N} \sum_{s=1}^{N-j} \sum_{b=0}^{\varrho(i)} (-1)^{s+\delta_{j,N}} z^{s+b-3}\left(\!\begin{array}{c}\!s\!+\!j\!-\!2\! \\ j-1\end{array}\!\right)\left(\!\begin{array}{c}\!\varrho(i)\! \\ \!b\!\end{array}\!\right) \times\,,
}
\\ \ \\
  \displaystyle{
\times \left(\! \frac{b \left(b-1 \right)}{2}\delta_{\varrho(i)-j,b+s-2}-b N z \delta_{\varrho(i)-j,b+s-1} \!\right) {E}_{i j}
    }
 \end{array}
 \eq
and
 \beq\label{w91}
 \begin{array}{c}
 \displaystyle{
B^{[1]}=\sum_{j=1}^{N}(-1)^{\delta_{j,N}}
\sum_{c=0}^{N-j} c ~  {E}_{\varrho^{-1}(j+c), \varrho^{-1}(\varrho(j)+c-1)}\,.
    }
 \end{array}
 \eq

 \underline{Matrix $m_{12}(0)$:}

Plugging $z=0$ into upper expressions we obtain:
 \beq\label{w92}
 \begin{array}{c}
 \displaystyle{
m_{12}(0)=A^{[0]}(0)\otimes B^{[1]}+A^{[1]}(0)\otimes B^{[0]}+A^{[2]}(0)\otimes 1_N+
    }
 \end{array}
 \eq
$$
+\sum_{i, j=1}^{N} {E}_{i j} \otimes\left\{-N (-1)^{\delta_{j,N}} \left[ \sum_{c=0}^{N-j} \Big((\varrho(i)+c) \delta_{\varrho(i)\leq j} -\varrho(i) \delta_{\varrho(i)\leq j+1} \Big){E}_{\varrho^{-1}(j+c), \varrho^{-1}(\varrho(i)+c-1)}+\right.\right.
$$
$$
\left.\left.+ \sum_{c=0}^{j-1} \Big(\varrho(i) \delta_{\varrho(i)>j+1} -(\varrho(i)-c-1) \delta_{\varrho(i)>j}  \Big) {E}_{\varrho^{-1}(j-c-1), \varrho^{-1}(\varrho(i)-c-2)} \right]\right\}
\,,
$$
where are $A^{[0]}(0)$, $A^{[1]}(0)$ and $A^{[2]}(0)$ are the following matrices:
 \beq\label{w93}
 \begin{array}{c}
 \displaystyle{
A^{[0]}(0)={E}_{N N}+ \sum_{i, j=1}^{N} (-1)^{\delta_{j,N}}
\delta_{\varrho(i),j-1} {E}_{i j}\,,
     }
 \end{array}
 \eq
 \beq\label{w94}
 \begin{array}{c}
 \displaystyle{
A^{[1]}(0)= \sum_{i, j=1}^{N} (-1)^{\delta_{j,N}}\left(\! \varrho(i)-N \!\right)\delta_{\varrho(i),j} {E}_{i j}
     }
 \end{array}
 \eq
and
 \beq\label{w95}
 \begin{array}{c}
 \displaystyle{
A^{[2]}(0)= \sum_{i, j=1}^{N} (-1)^{\delta_{j,N}}\left(\! \frac{\varrho(i) \left(\varrho(i)-1 \right)}{2}-N \varrho(i) \!\right)\delta_{\varrho(i),1+j} {E}_{i j}\,,
     }
 \end{array}
 \eq

 \underline{Matrix $r_{12}^{(0)}$:}

 \noindent The matrix $r_{12}^{(0)}$ is the coefficient in the expansion
 \beq\label{w96}
 \begin{array}{c}
 \displaystyle{
 r_{12}(z)=\frac{P_{12}}{z}+r_{12}^{(0)}+O(z)\,.
     }
 \end{array}
 \eq
It is as follows:
 \beq\label{w97}
 \begin{array}{c}
 \displaystyle{
r_{12}^{(0)}=A^{[0]}(0)\otimes B^{[0]}+A^{[1]}(0)\otimes 1_N+ \sum_{i, j=1}^{N} {E}_{i j} \otimes\Big\{\varrho(i)
{E}_{j, \varrho^{-1}(\varrho(i)-1)}-
     }
 \end{array}
 \eq
$$
-\sum_{\gamma=0}^{\varrho(i)} (-1)^{\varrho(j)+N}\delta_{\gamma + N,j}(N-j)\left(\begin{array}{c}\varrho(i) \\ \gamma\end{array}\right)\left(\begin{array}{c}N \\ j-1\end{array}\right) {E}_{N, \varrho^{-1}(\varrho(i)-\gamma)}-(-1)^{\delta_{j,N}} j {E}_{\varrho^{-1}(j), i}+
$$
$$
+N (-1)^{\delta_{j,N}}
\Big[\delta_{\varrho(i)\leq j}\sum_{c=0}^{N-j} {E}_{\varrho^{-1}(j+c), \varrho^{-1}(\varrho(i)+c)}
-\delta_{\varrho(i)>j} \sum_{c=0}^{j-1} {E}_{\varrho^{-1}(j-c-1), \varrho^{-1}(\varrho(i)-c-1)}\Big]\Big\}
\,.
$$

\subsection*{Acknowledgments}


This work was supported by the Russian Science Foundation under grant no.19-11-00062,\\ https://rscf.ru/en/project/19-11-00062/.


\begin{small}

\end{small}

\end{document}